\documentclass[a4paper,12pt]{article}

\usepackage{amsmath,amssymb,algorithmic}
\usepackage{latexsym}
\usepackage[latin1]{inputenc}
\usepackage[T1]{fontenc}
\usepackage{textcomp}
\usepackage{times}

\usepackage[pdftex]{graphicx}
\usepackage[frenchb]{babel}
\usepackage{babelbib}
\usepackage{hyperref}
\hypersetup{
colorlinks=true, 
citecolor=blue, %
pdftoolbar=false,
bookmarksopen=true,
breaklinks=true, 
urlcolor=blue, 
linkcolor=blue, 
}

\author{Philippe Besse\thanks{Université de Toulouse -- INSA, Institut de Mathématiques, UMR CNRS 5219} \and Béatrice Laurent\thanks{Université de Toulouse -- INSA, Institut de Mathématiques, UMR CNRS 5219}}
\title{\sc De Statisticien à Data Scientist \\ {\normalsize Développements pédagogiques à l'INSA de Toulouse}}

\begin{document}
\sloppy
\DeclareGraphicsExtensions{.pdf,.jpg}
\maketitle

\begin{quote}
{\bf Résumé}: 
Selon un \href{http://europa.eu/rapid/press-release_IP-14-1129_fr.htm}{rapport  récent} de la  commission européenne, le monde génère chaque minute 1,7 millions de milliards d'octets de données, soit l'équivalent de 360 000 DVD, et les entreprises qui bâtissent leur processus décisionnels en exploitant ces données accroissent leur productivité. Le traitement et la valorisation de données massives a des conséquence en matière d'emploi pour les diplômés des filières statistiques.  Quelles compétences nouvelles les étudiants formés en statistique doivent-ils acquérir devenir des \emph{scientifiques des données}?  Comment faire évoluer les formations pour permettre aux futurs diplômés  de s'adapter  aux évolutions rapides dans ce domaine, sans pour autant négliger les métiers traditionnels et le socle fondamental  et pérenne de la formation?  Après nous être interrogés sur la notion de données massives et l'émergence d'une "nouvelle" science : \emph{la science des données}, nous présenterons les évolutions  en cours dans la formation d'ingénieurs en Génie Mathématique et Modélisation à l'INSA de Toulouse. 
{\bf Mots-clefs}: Science des données; Statistique; données massives; enseignement

{\bf Abstract}: 
According to a \href {http://europa.eu/rapid/press-release_IP-14-1129_fr.htm}{recent report} from the European Commission, the world generates every minute 1.7 million  of billions  of data bytes,  the equivalent of 360,000 DVDs, and companies that build their decision-making processes by exploiting these data increase their productivity. The treatment and valorization of massive data has consequences on the employment of graduate students in statistics.  Which additional skills do students trained in statistics  need to acquire to become \emph {data scientists ? }  How to evolve training  so that future graduates   can  adapt to rapid changes in this area, without neglecting traditional jobs and the fundamental and lasting foundation for the training? After considering   the notion of big data and  questioning  the  emergence of a "new" science: \emph {Data  Science}, we present the current developments in the training of engineers in  Mathematical and Modeling at INSA Toulouse.

{\bf Keywords}: Data Science; Statistics; big data; teaching.
\end{quote}

\section*{Introduction}
Le déluge ou le {\it tsunami} massif des données et ses conséquences en matière d'emploi des étudiants formés à la Statistique, bouscule le paysage académique. Beaucoup de Masters ont été créés ces deux dernières années avec un intitulé {\it Sciences des Données}. L'\href{http://www.insa-toulouse.fr/fr/index.html}{INSA de Toulouse}, qui intègre une spécialité d'ingénieurs \href{http://www.math.insa-toulouse.fr/fr/index.html}{\it Génie Mathématique et Modélisation} (GMM) incluant une orientation {\it Modèles et Méthodes Statistiques} (MMS) n'échappe pas à ce mouvement de fond. L'objectif de cet article est de décrire la stratégie adoptée et les développements pédagogiques mis en place pour accompagner cette évolution à très court terme puis à moyen terme avec le renouvellement à venir de l'accréditation de la spécialité GMM par la Commission du Titre d'Ingénieur courant 2016 pour la rentrée 2017. 

Le département de Mathématiques de l'INSAT forme des statisticiens qui intègre avec succès le monde professionnel, comment former des \emph{scientifiques des données}? Plus précisément, nous formons  depuis la promotion 2007 des étudiants  aux méthodes récentes de modélisation et apprentissage statistiques; depuis la promotion 2013, ceux-ci sont embauchés comme \emph{data scientist}. Nous formions donc à des métiers\footnote{\`A l'insu de notre plein gré.} avant que ceux-ci n'existent. Plus sérieusement, comment mieux former les étudiants à la réalité actuelle de ces emplois?

Ces changements touchent bien sûr les contenus des programmes mais impactent également nos pratiques pédagogiques. La première section de cet article décrit quelques  enjeux en lien avec le traitement des données massives, la deuxième s'interroge sur l'émer\-gence d'une "nouvelle science". La troisième section résume succinctement l'environnement disciplinaire des étudiants de GMM MMS et les nouveaux objectifs visés. La quatrième section  décrit les adaptations de programme, principalement à court terme, leurs objectifs pédagogiques et les nouvelles ressources associées qui sont disponibles sur le site \href{http://wikistat.fr/}{\tt wikistat.fr}.

\section{Quelques enjeux des données massives}
\subsection{Disparité}
Malgré une apparente unité, la \emph{Science des données} recouvre des réalités très différentes en fonction de leurs origines et modes de production. La partie la plus visible concerne les principaux acteurs d'internet (GAFA ou Google, Apple, Facebook, Amazon), ce sont les premiers producteurs, analyseurs de données massives et sans doute les financeurs majeurs de la recherche dans ce domaine, devant les financements publics. Le contexte est complètement différent lorsqu'il s'agit de données publiques ou \emph{open data}, de données scientifiques comme par exemple en Astronomie, Biologie, ou encore de données industrielles. Les problèmes de confidentialité, de sécurité, d'archivage, de besoins de calcul, d'algorithmes... se posent de façon radicalement différentes. Certes, beaucoup des méthodes mathématiques et statistiques utilisées sont transférables d'un domaine à l'autre, mais le contexte technologique de mise en \oe uvre change considérablement les choix en présence, notamment dans la façon d'architecturer, paralléliser les stockages et calculs, donc les choix de méthodes et d'algorithmes et finalement les compétences requises. 

\subsection{Réellement massives?}
La vogue de l'appellation \emph{big data} a eu des conséquences immédiates sur la communication. Tout problème maintenant classique de fouille de données  (\emph{data mining}), par exemple en marketing quantitatif ou gestion de la relation client, est devenu de la \emph{Science des données} alors que celles-ci sont loin d'être massives. Pour ajuster le discours, les données deviennent  \emph{smart} ou encore l'aspect "\emph{big}" de la communication concerne les résultats commerciaux attendus ou promis par l'exploitation des données, plus que leur volume.

\subsubsection*{Seuils technologiques}
D'autres domaines produisent effectivement des données massives confrontées à deux seuils technologiques liés au volume. Le premier concerne la mémoire interne (RAM) de l'ordinateur. La majorité des logiciels scientifiques (R, Matlab, Scikit-learn de Python...) nécessitent le chargement en mémoire de l'ensemble des données pour leur traitement. Quels choix opérer en cas d'impossibilité: changer d'ordinateur (ce peut être la meilleure solution), échantillonner, répartir les données et les calculs sur un réseau ou \emph{cluster} d'ordinateurs? Ce dernier choix rejoint le deuxième seuil technologique lorsque leur volume dépasse la capacité de stockage d'un seul ordinateur (disons actuellement quelques tera-octets) ou encore si le temps de réponse nécessite un traitement parallélisé des données. En conséquence, la problématique: \emph{données massives} et ses implications méthodologiques, algorithmiques, émergent réellement à partir du moment où les données sont \emph{distribuées} sur plusieurs ordinateurs. Dans le cas contraire, les méthodes statistiques et algorithmes usuels d'apprentissage restent opérants. 

Bien entendu les questions de \emph{vélocité} de l'acquisition, associées à des prises de décision en ligne ou en temps réel, la complexité ou la \emph{variété} (images graphes, signaux) des données, soulèvent d'autres questions plus fondamentales de représentation (objet mathématique) indépendamment du volume considéré. Ces questions apparues bien avant la "naissance" de la science des données concernent alors plus des sujets de recherche (thèse) qu'une formation de niveau Master. 

\subsubsection*{Préparation des données}
Plus précisément et encore plus concrètement en amont, la phase de traitement la plus importante, en terme d'implication humaine et de trafic de données, est souvent celle d'extraction, préparation. Cette étape, dite aussi de \emph{data munging} ou \emph{wrangling}: vérification, imputation, transformation, sélection des variables (\emph{features}), est unidimensionnelle (au plus bi), facilement  distribuable donc parallélisable dans un langage comme Python éven\-tuellement en connexion avec des technologies spécifiques \href{https://hadoop.apache.org/}{Hadoop} et \href{http://spark.apache.org/}{Spark}. La préparation des données met en \oe uvre des méthodes de niveau très élémentaire (résumé statistique des distributions) à très élaborée (recalage de courbes ou \emph{time warping}, d'images...), elle requiert souvent des compétences "métiers" du domaine d'application et est, de toute façon, fondamentale pour la qualité des résultats. Elle est malheureusement souvent occultée lors des présentations commerciales des technologies en concurrence pour données massives qui se contentent de données "jouet" (cf. les inévitables \href{https://archive.ics.uci.edu/ml/datasets/Iris}{iris de Fisher}). Une fois cette étape réalisée, les données extraites et mises en forme sont finalement rarement volumineuses et les algorithmes, programmes classiques d'exploration, modélisation ou apprentissage sont utilisables sur un poste personnel ou un gros serveur capable de les charger en mémoire. 

\subsubsection*{\emph{Scalability} des méthodes et algorithmes}
Dans son exposé lors des Journées de Statistique de Lille (2015), David Bessis a été très clair sur ce point. Les méthodes (factorisation de matrices non-négatives ou NMF) que sa société (\href{https://www.tinyclues.com/}{Tinyclues})  met en \oe uvre pour des systèmes de recommandation de sites de commerce en ligne s'exécutent en mémoire sur un serveur d'\href{https://aws.amazon.com/fr/}{Amazon Web Service}. Certaines situations peuvent néanmoins nécessiter l'exécution d'algorithmes, par exemple d'apprentissage supervisé ou non, sur des données physiquement distribuées. Ce n'est guère que dans ce cas que des technologies et librairies dédiées deviennent incontournables, elles mettent en \oe uvre des méthodes susceptibles d'être décomposées en phases successives de \emph{MapReduce}, on dit qu'elles sont alors \emph{échelonnables} comme traduction de \emph{scalable}. C'est à ce jour la librairie \href{http://spark.apache.org/docs/latest/mllib-guide.html}{\tt MLlib} de \href{http://spark.apache.org/}{\it Spark} qui semble la plus prometteuse ou tout du moins celle la plus couramment citée. Elle utilise le principe introduit dans Spark de {\it base de donnée résiliente} (Zaharia et al. 2012) afin de rendre efficace des algorithmes itératifs en minimisant les accès disques sur le  système \href{https://hadoop.apache.org/}{\it Hadoop} de gestion de fichiers partagés. 

\subsection{Réalité ou fantasmes?}
La croissance exponentielle de la production de données et leurs trois V (volume, variété, vélocité)  ont été et sont toujours très médiatisés pour glorifier leurs conséquences en matière de croissance économique, nombre d'emplois, avancées scientifiques. En contrepartie, mais rarement dans le même article, la même emphase est déployée pour mettre en exergue les risques associés: éthiques, juridiques, sécuritaires ou politiques. Cet article n'est pas le lieu pour aborder ces points  mais il reste important de sensibiliser, voire d'alerter, les étudiants sur certains aspects. Donnons deux exemples: 
\begin{itemize}
\item Confusion, suite à l'affaire Snowden,  entre la face cachée ou \emph{big brother} du stockage massif des messages personnels (NSA), d'une part et la prévision de comportements ou de goûts personnels vendue à des agences de publicité d'autre part. De plus, le statisticien sait bien que le volume, en terme de nombre d'individus, permet d'atteindre une grande précision pour la prévision de comportements moyens (loi des grands nombres) mais une part irréductible d'aléa, c'est heureux, reste attaché à la prévision d'un comportement individuel. Que penser des affirmations du site \href{http://applymagicsauce.com/you.html}{Apply Magic Sauce}\footnote{Cité dans un \href{http://tempsreel.nouvelobs.com/les-internets/20150903.OBS5238/facebook-voila-ce-que-vos-like-revelent-de-votre-personnalite.html}{article du Nouvel Observateur}.} qui "révèle votre personnalité" à partir de ce que vous aimez (vos \emph{Like}) de \emph{Facebook} ? 
\item Depuis quelques années aux USA et plus récemment en France, des assurances autos \footnote{Cité dans un \href{http://tempsreel.nouvelobs.com/economie/20150902.OBS5154/assurances-vous-paierez-moins-si-vous-conduisez-bien.html}{article du Nouvel Observateur}} proposent l'installation d'un boîtier embarqué pour évaluer la qualité de conduite afin de personnaliser chaque mois le tarif de la prime : \emph{payez comme vous conduisez}. Quelles sont les implications éthiques d'un tel procédé s'il se généralise, par exemple, à l'assurance maladie ou à celle d'un prêt immobilier ? C'est au législateur de réguler l'équilibre entre une segmentation de plus en plus fine de la clientèle donc des tarifs, d'une part et d'autre part la solidarité ou mutualisation des risques qui repose sur le principe d'\emph{asymétrie de l'information} au profit, en principe et jusqu'à présent, de l'assuré.
\end{itemize}

\section{Nouvelle science?}
\subsection{Nouveau paradigme?}
Présente depuis plusieurs siècles dans toute analyse de données, que manque-t-il à la Statistique pour se voir suppléer par la \emph{Science des données}? Quel nouveau paradigme serait récemment apparu pour justifier de la création de "nouveaux" métiers? Depuis le développement de la Statistique mathématique du milieu du siècle dernier, cette discipline a évolué parallèlement et inévitablement aux innovations technologiques et donc au volume des données traitées. Ignorer ces développements,  c'est penser que cette discipline se limite à une expérience planifiée, des calculs de moyennes écarts-types, une droite de régression et un ou deux tests (Student, Fisher). 

Ce qui est nouveau c'est de toute évidence une forme de digitalisation ou "datafication" massive du quotidien. Photos, messages, documents, navigation internet, contacts, agenda, géolocalisation... , tout se numérise, s'archive et devient accessible à l'analyse; ce phénomène ne peut que s'accentuer avec la connexion de toujours plus d'objets personnels. Mais qu'en est-il du point de vue méthodologique pour analyser, exploiter, valoriser ces données.
\subsubsection*{Données préalables à l'analyse}
Un changement notable est apparu avec l'avènement du \emph{data mining} dans les années 90. Premier changement de paradigme, l'acquisition des données n'est plus \emph{planifiée}. Il s'agit de valoriser celles stockées, principalement pour des raisons comptables, afin d'optimiser la gestion de la relation client (GRC) dans les grandes sociétés tertiaires: banque, assurance, téléphonie, VPC... Le statisticien devient "prospecteur de données". L'ouverture massive de sites de vente en ligne fait actuellement exploser ces besoins en marketing quantitatif sans nécessairement renouveler les méthodes. Certes, les systèmes de recommandation en temps réel avec décision séquentielle sont massivement apparus avec le commerce en ligne mais ils remettent surtout au goût du jour les algorithmes de bandits (Thomson, 1933) illustrés par le succès commercial de \href{http://www.criteo.com/fr}{Criteo} et  suscitent évidemment de nouvelles recherches (cf. par exemple Cappé et al. (2013)) mais post Master.

\subsubsection*{$p>>n$}
Le développement des biotechnologies depuis le début du siècle facilite et popularise la production de masses de données conséquentes notamment avec les techniques récentes de séquençage et autres omiques. Le deuxième changement de paradigme est ainsi introduit par l'accroissement considérable du nombre $p$ d'informations (expression de gènes, de protéines...) disponibles par échantillon biologique alors que la taille $n$ de ces échantillons reste modeste. Analyser plusieurs dizaines de milliers à millions de variables pour quelques dizaines ou au plus centaines d'individus, c'est plus d'indétermination que d'information. Après la correction des tests multiples (Benjamini \& Hochberg), les méthodes statistiques ont été adaptées à cette situation par l'introduction systématique de phases de sélection de variables, par exemple par la prise en compte d'une contrainte de pénalisation en norme $L_1$ (cf. Lê Cao et al.  2011 pour la régression PLS). Le statisticien est devenu "bioinformaticien" ou au moins publie dans des revues de Bioinformatique.

\subsubsection*{Erreur d'optimisation}
Les problèmes méthodologiques soulevés par la variété des données ou leur vélocité motivent de nouveaux développements mathématiques et statistiques (algorithmes adaptatifs, décision séquentielle, courbes, surfaces, graphes, texte en langage naturel...) mais sans rupture dans la prise en compte de toujours plus de complexité. En revanche, le volume, s'il est réel, par rapport à des capacités de calcul et de stockage limitées, introduit un nouveau problème ou plus précisément un nouveau terme d'erreur à minimiser; c'est le troisième changement de paradigme.

Au c\oe ur d'un \href{http://wikistat.fr/pdf/st-m-app-risque-estim.pdf}{cours d'apprentissage statistique} ou machine, il y a le contrôle du sur-apprentissage ou l'optimisation de la complexité d'un modèle pour minimiser l'erreur de prévision (estimée généralement par validation croisée) en équilibrant au mieux biais (erreur d'approximation) et variance (erreur d'estimation) des prévisions car si le biais décroît avec la complexité d'un modèle, la variance  augmente et peut exploser avec des problèmes mal conditionnés et le sur-apprentissage.

Le troisième et nouveau terme d'erreur d'optimisation,  peut s'illustrer sur un exemple simple: quelle est la meilleure stratégie? C'est-à-dire celle conduisant à la plus petite erreur de prévision:
\begin{itemize}
\item \'Echantillonner les données pour en réduire la taille (inférieure à celle de la mémoire) et se ramener à un déroulement classique d'optimisation biais / variance avec un grand choix de méthodes utilisables tout en introduisant une erreur due à la réduction de l'ensemble d'apprentissage.
\item Estimer un modèle sur toutes les données, alors nécessairement distribuées, en utilisant par exemple un environnement \href{https://hadoop.apache.org/}{\it Hadoop} / \href{http://spark.apache.org/}{\it Spark} et la librairie \href{http://spark.apache.org/docs/latest/mllib-guide.html}{\tt MLlib}. Le choix de méthode disponible devient très réduit, seules celles passant à l'échelle (\emph{scalable}) sont disponibles et les procédures d'optimisation de leurs paramètres de complexité souvent négligées car fort coûteuses en temps de calcul.
\end{itemize}
Les ressources (mémoire, temps) sont contraintes, comment minimiser le terme d'erreur d'optimisation dû, soit à l'échantillonnage, soit à la limitation des calculs ou méthodes utilisables sur une base d'apprentissage très volumineuse et distribuée? De façon globale, une fonction coût à minimiser peut être définie s'il est possible d'évaluer le coût induit par des erreurs de prévision, à mettre en balance avec le coût du calcul pour l'apprentissage sur des serveurs proposant ce service.

\subsubsection*{Exhaustivité \emph{vs.} échantillonnage}
L'exemple précédent soulève une nouvelle remarque ou de nouvelles questions. Jusqu'à la fin du XIXème siècle, les statistiques étaient principalement exhaustives. C'est avant la 2ème guerre, notamment avec les travaux de Neyman (1936), que s'est développée la théorie des sondages, donc de l'échantillonnage, puis que sont apparus les premiers instituts ({\it e.g.} Gallup) pour étudier, par exemple, le comportement des consommateurs. Nouvelle évolution, l'avènement des données massives rend accessible l'observation et, pourquoi pas, le retour vers le futur avec à nouveau l'\emph{analyse exhaustive} d'une population. Certes, en l'absence de contrainte de calcul comme vue précédemment, un tel traitement élimine l'erreur due à l'échantillonnage et la contribution d'un statisticien.

 Malheureusement, les données exhaustives récoltées sont-elles bien représentatives de la réalité? Ou encore, beaucoup de données signifie-t-il beaucoup d'informations ou une information utile? L'exemple des mesures d'audience réalisées par \href{http://www.mediametrie.fr/}{Médiamétrie} illustre bien ces questions. Certes les fournisseurs d'accès savent précisément quelle chaîne de télévision est diffusée par la  boîte de tous leurs abonnés mais cette dernière ne sait pas qui est devant la télévision ou même si seulement la télévision est allumée; un site internet enregistre tous les clics de tous les visiteurs mais quel est le profil d'un visiteur (60\% sont en moyenne des robots)? C'est la raison pour laquelle \href{http://www.mediametrie.fr/}{Médiamétrie} propose un nouveau modèle, nécessairement statistique, de   \href{http://www.mediametrie.fr/innovation/pages/la-mesure-d-audience-hybride-de-l-internet-fixe.php?p=7,112,116&page=92}{mesure d'audience hybride} de l'internet fixe. Cette mesure utilise les fichiers log de connexions pour redresser les données individuelles d'un panel classique d'internautes bien identifiés.

\subsection{Quelle économie?}
\subsubsection*{Pressions socioéconomiques}
 Il y a 20 ans (cf. par exemple Besse et al. 2001) naissait le \emph{Data Mining} déjà promu comme une nouvelle discipline mais qui se résume principalement à l'exploitation de suites logicielles commerciales comme \href{http://spss-clementine.software.informer.com/}{IBM SPSS Clementine}, \href{http://support.sas.com/software/products/miner/}{SAS Enterprise Miner}, \href{http://www.coheris.com/produits/analytics/logiciel-data-mining/}{SPAD},  ou libres comme \href{https://www.knime.org/knime}{Knime}, \href{http://eric.univ-lyon2.fr/~ricco/tanagra/}{Tanagra}, \href{http://www.cs.waikato.ac.nz/ml/weka/}{Weka}...  interfaçant  amicalement\footnote{Interface graphique \emph{user friendly}.} des outils de gestion de bases de données avec des algorithmes d'exploration, modélisation statistique (régression gaussienne, logistique...)  et d'apprentissage machine (réseaux de neurones...) par ailleurs bien connus. L'histoire se répète avec une emphase proportionnelle au facteur d'échelle multiplicatif du volume des données (méga, giga, téra...) et la mise en place de sites ou plateformes de service.
 
L'émergence massive d'un phénomène médiatique rend nécessaire, pour des raisons de stratégie de communication commerciale plus que scientifiques, le renouveau des étiquettes. Au revoir la \emph{Statistique}, avec ses spécialistes académiques certifiés pour ne pas dire passéistes issus du siècle dernier,  bonjour la \emph{Science des Données}, thématique suffisamment vaste et aux frontières mouvantes sur laquelle tout le monde ou presque peut s'exprimer, eldorado, des nouveaux conquistadores de parts de marché. 

\subsubsection*{Nouveau modèle économique}
Le paysage\footnote{Consulter la \href{http://mattturck.com/2012/06/29/a-chart-of-the-big-data-ecosystem/}{carte} de Matt Turk} ou plutôt la jungle économique et industrielle des technologies de l'information est en profonde mutation. Des acteurs majeurs: IBM, Hewlett Packard, Oracle, Microsoft, SAS Institute..., sont fortement ébranlés dans leurs c\oe urs de métier historiques et se reconvertissent avec plus ou moins de succès et d'efficacité pour suivre le mouvement: compatibilité avec le foisonnement de la production de code sous une licence {\it open source} comme celle des projets de la \href{http://www.apache.org/}{fondation Apache}, investissement massif vers les activités de services à forte valeur ajoutée plutôt que la production de matériels et même de logiciels. Citons simplement l'exemple d'Amazon\footnote{Source : \href{http://www.lemondeinformatique.fr/actualites/lire-aws-conserve-sa-place-de-numero-1-mondial-du-cloud-60113.html}{ICT Journal}}: AWS (\href{https://aws.amazon.com/fr/}{Amazon Web Service}) ne représentait que 5,3\% du chiffre d'affaire d'Amazon au premier semestre 2014, il est passé à 7\% avec une croissance de 49\% en un an. AWS reste en tête de ce marché (30\%)\footnote{Source:\href{http://www.lemondeinformatique.fr/actualites/lire-aws-conserve-sa-place-de-numero-1-mondial-du-cloud-60113.html}{Le Monde de l'Informatique}} devant \href{http://www.microsoft.com/en-us/server-cloud/products/analytics-platform-system/}{Microsoft} et sa \href{http://azure.microsoft.com/fr-fr/services/machine-learning/}{plateforme}, \href{http://www.ibm.com/cloud-computing/us/en/}{IBM} et même \href{https://cloud.google.com/}{Google}.

Le modèle économique est profondément remanié. Les équipements trop chers sont loués, les langages et logiciels sont libres mais les concepts et méthodes trop complexes à assimiler et mettre en \oe uvre sont des services (formations, plateformes) monnayables. Plus généralement, tout un ensemble de ces services et une nomenclature associée se développent avec l'industrialisation, la commercialisation du \emph{cloud computing}: \emph{software as a service} (SaaS) , \emph{infrastructure as a service} (IaaS), \emph{platform as a service} (PaaS), \emph{desktop as a service} (DaaS), \emph{backend as a service} (BaaS), \emph{information technology management as a service} (ITMaaS). Cet article n'est pas le lieu de développer ces points. Citons seulement quelques entreprises surfant sur la vague des nouvelles  technologies: \href{https://www.enthought.com/}{Enthought} (Canopy) et \href{http://continuum.io/}{Continuum analytics} (Anaconda) proposent des distributions libres de Python et, c'est important, faciles à installer ainsi que des versions plus élaborées payantes et de la formation associée. Les créateurs (Zaharia et al. 2012) de \href{http://spark.apache.org/}{Spark} ont fondé \href{https://databricks.com/}{databricks}: \emph{Data science made easy, from ingest to production}, pour principalement vendre de la formation et une certification. Trevor Hastie et Ron Tibshirani conseillent \href{http://h2o.ai/about/}{Oxdata} qui développe une forme d'interface: \href{http://h2o.ai/#/}{H20} entre \href{https://cran.r-project.org/}{R} et \href{https://hadoop.apache.org/}{Hadoop} avec des algorithmes spécifiques dont une version édulcorée d'apprentissage profond (\emph{deep learning}).

\subsection{Quelle Mathématique?}

\subsubsection*{Nouvelles méthodes?}
Qu'est-ce qui justifierait l'appellation d'une nouvelle science dite des données? Le troisième terme d'erreur d'optimisation décrit ci-dessus, conséquence directe du changement d'échelle en volume, serait-il à l'origine du changement d'étiquette? L'association systématique faite entre "science des données" et technologies de stockage distribuées (\href{https://hadoop.apache.org/}{Hadoop}) le laisserait penser alors que les aspects les plus fondamentaux sont finalement peu impactés; on assiste plutôt à un retour en arrière (cf. Besse et al. 2014), une remise à jour de méthodes relativement anciennes comme $k$-\emph{means} (Mac Queen, 1967), régression logistique et plus récentes (forêts aléatoires de Breiman; 2001), dont les algorithmes ont de bonnes propriétés (\emph{scalable}) pour être distribuées, parallélisées. 

L'adaptation algorithmique qui en résulte et le nouveau terme d'erreur à prendre en compte font nécessairement évoluer la recherche dans les disciplines concernées mais peut-on parler de nouvelle science si ce n'est que statisticiens, mathématiciens (optimisation) et informaticiens (bases de données, architectures...) sont clairement obligés de collaborer. 

\subsubsection*{Recherche fondamentale}
Une nouvelle Science se devrait d'ouvrir et explorer de nouveaux champs fondamentaux de recherche. Le \emph{buzz} médiatique bouscule les frontières classiques interdisciplinaires. Ce ne peut être que très positif, surtout en France où celles-ci sont  rendues peu perméables par les structures de nos instances nationales de recrutement et promotion (\href{http://www.cpcnu.fr/listes-des-sections-cnu}{Comité National des Universités}, \href{http://www.cnrs.fr/comitenational/sections/intitsec.php}{Comité national de la Recherche Scientifique}). Le CNRS vient\footnote{Assemblée générale constitutive des 24 et 25 juin 2015.} de labéliser le nouveau Groupement de Recherche \href{http://www.madics.fr/}{MaDICS} (Masse de Données, Information et Connaissance en Sciences). 
\begin{quote}
\href{http://www.madics.fr/}{MaDICS} propose un écosystème \emph{Masses de données scientifiques} afin de promouvoir et animer des activités de recherche interdisciplinaires\footnote{Difficile d'avoir des chiffres précis mais la très grande majorité des présents à cette assemblée émargeaient aux sections 6/7 du CNRS ou 27 du CNU!} positionnées dans un continuum \emph{des données aux connaissances et à la prise de décision} dont le point de départ sont les masses de données en Sciences. MaDICS est également un forum d'échanges entre scientifiques et acteurs économiques confrontés aux problèmes du "big data" et des Sciences des données, un instrument de prospective et un lieu d'accompagnement des jeunes chercheurs dans les domaines concernés.
\end{quote}
Néanmoins les avancées majeures, dynamisées par les nouveaux défis, restent profondément disciplinaires en lien avec des fondamentaux tant en Mathématiques qu'en Informatique, sans faire naître une nouvelle science. Ainsi, le concours \href{http://www.netflixprize.com/}{Netflix} à un million de dollars a suscité une vague importante de publications à la suite des travaux de Candes et Tao (2009) en optimisation convexe pour la complétion de très grandes matrices creuses non négatives. En lien avec le troisième terme d'erreur, ce sont  généralement et dans la plupart des colloques sur le sujet, des problèmes d'optimisation qui sont approfondis d'un point de vue théorique, mathématique (gradient stochastique...) et algorithmique pour optimiser les performances des librairies et  les adapter à des architectures spécifiques.

Le point de vue adopté dans cet article est évidemment celui de statisticiens entraînés à l'analyse de données de toutes origines; le spécialiste d'une autre discipline peut en adopter un autre. Le travail du biologiste est largement impacté par le développement récent des  biotechnologies dont le séquençage à haut débit. Les analyses associées ont profité et aussi motivé le développement des méthodes d'apprentissage en très grande dimension ces dernières années. La Physique fait de même. Lorsqu'un système est trop complexe (turbulences) pour l'explicitation d'un modèle analytique ou son calcul explicite (Schrödinger), un (méta)modèle d'apprentissage statistique \emph{boîte noire} non interprétatif mais prédictif peut s'avérer utile comme par exemple pour défier le \href{https://higgsml.lal.in2p3.fr/}{boson de Higgs}; un outil parmi d'autres, qui ne  remplacera pas la Physique.

\subsection{En résumé}
Une visite au site présentant les services d'\href{https://aws.amazon.com/fr/machine-learning/details/}{\emph{Amazon WS Machine learning}} illustre  bien ce débat. Il y est affirmé que les outils développés permettent "\emph{d'utiliser une technologie d'apprentissage-machine puissante sans avoir besoin de maîtriser les algorithmes et techniques de l'apprentissage-machine}" (\emph{sic}!). La promotion commerciale de certains logiciels de prospection de données promettait déjà il y a quinze ans l'extraction de pépites (\emph{nuggets}) sans se salir les mains. Au jour (septembre 2015) de la visite du site,  AWS déploie et commercialise des moyens de stockage et calcul impressionnants mais plus que restreints\footnote{Source: \href{https://docs.aws.amazon.com/machine-learning/latest/dg/learning-algorithm.html}{Mode d'emploi} d'AWS.} d'un point de vue méthodologique : le choix de méthode est strictement limité au modèle linéaire (général): régressions gaussienne, binomiale ou logistique, logistique polytomique. La prise en compte de possibles interactions se fait "à la main" en créant de nouvelles variables par produit cartésien de variables qualitatives, le réglage de la pénalisation $L_1$ sensée opérer la sélection de variables est laissé au choix de l'utilisateur sans automatisation de l'optimisation et avec une valeur par défaut du paramètre à 0; le site renvoie sur \href{https://en.wikipedia.org/wiki/Regularization_%28mathematics%29}{Wikipedia} pour les explications.

Vue sous cet angle, la "Science des Données" devient un enrobage publicitaire (\emph{packaging}) pour masquer des approches statistiques limitées (\href{https://aws.amazon.com/fr/machine-learning/}{AWS}), ou élaborées (\href{https://databricks.com/}{databricks}), exécutées dans des environnements technologiques sophistiqués dans le but de louer de l'espace disque et vendre des heures de calcul ainsi que  de la formation. 

Dans le cas d'objectifs de marketing pour le calcul de score d'appétence ou d'attrition à partir de données massiques, on peut imaginer que la rapidité de mise en \oe uvre, la réduction des coûts de personnels, incitent à l'utilisation "presse bouton" de plateformes dédiées plutôt qu'à des analyses sur mesure élaborées avec les compétences adaptées. Quelques points d'erreur de prévision en plus n'ont pas de conséquences redoutables dans ce domaine. Néanmoins, compte tenu des investissements consentis en amont, cette stratégie n'est sans doute pas optimale, elle est même inacceptable dans d'autres secteurs d'application: recherche, Santé publique, industrie.

\section{L'enseignement de la Statistique à l'INSAT}
\subsection{Orientation GMM MMS}
Comme les autres écoles du  \href{http://www.groupe-insa.fr/}{groupe INSA}, celui de \href{http://www.insa-toulouse.fr/fr/formation/ingenieur.html}{Toulouse} recrute, principalement après le bac, pour 5 ans de formation organisée, c'est original (1+2+2) à Toulouse\footnote{D'autres recrutements sont prévus sur dossier aux niveaux L2, L3, M1.}, avec une année commune, deux années de pré-orientation et deux de spécialisation. La pré-orientation Modélisation Informatique et Communication (\href{http://www.insa-toulouse.fr/fr/formation/ingenieur/preorientations/po_mic_modelisation_informatique_et_communication.html}{MIC}) conduit  à deux spécialités d'ingénieur dont celle de Génie Mathématique et Modélisation (\href{http://www.insa-toulouse.fr/fr/formation/ingenieur/specialites/gm.html}{GMM}) au cours de laquelle les étudiants ont le choix entre deux orientations Méthodes et Modèles Numériques d'une part, Statistiques d'autre part. Cette dernière filière concerne chaque année une trentaine d'étudiants qui suivent une quatrième année (\href{http://odf.insa-toulouse.fr/fr/formation/formation_ingenieur/offre_de_formation/cursus_ingenieur/sciences_et_technologies/ingenieur_genie_mathematique_et_modelisation/quatrieme_annee_genie_mathematique_et_modelisation.html}{programme détaillé})avec les contenus relativement classiques d'un Master de Statistique. Les étudiants ont ensuite le choix, en cinquième  année (\href{http://odf.insa-toulouse.fr/fr/formation/formation_ingenieur/offre_de_formation/cursus_ingenieur/sciences_et_technologies/ingenieur_genie_mathematique_et_modelisation/cinquieme_annee_genie_mathematique_et_modelisation.html}{programme détaillé}), entre deux groupes d'unités de formation (UF) en plus d'un tronc commun. Deux UFs sont consacrées à l'ingénierie mathématique pour la Finance et deux aux applications industrielles de la Statistique (risque, incertitude analyse de sensibilité, planification expérimentale). Le tronc commun regroupe un projet tutoré "Recherche Innovation" et un cours assez volumineux d'\href{http://www.math.univ-toulouse.fr/~besse/Wikistat/pdf/st-m-app-intro.pdf}{apprentissage statistique} (60h), de la modélisation dite avancée (durée de vie, statistique spatiale) et des "humanités".
\subsection{Statistique et employabilité}
La Statistique est évidemment présente comme discipline de service dans les huit spécialités représentées à l'INSAT mais également comme une spécialité spécifique au sein de celle de modélisation mathématique.  Les étudiants ainsi formés trouvent très rapidement un emploi à leur sortie dans des secteurs très divers couvrant actuellement les applications financières (analyse de risque...), l'Actuariat (moyennant une année supplémentaire pour obtenir le titre), l'industrie (le plus souvent avec une thèse CIFRE), la Biostatistique (généralement pour faire une thèse) et la fouille des données ({\it data mining}) avec des applications par exemple en marketing quantitatif ou gestion de la relation client (GRC).

Depuis la promotion 2013, les étudiants, qui apprennent aussi  à communiquer, se désignent dans leur CV comme étant des {\it Data Scientists} et sont embauchés sur de tels intitulés de postes. L'adaptation à cette nouvelle réalité de l'offre d'emploi fut indispensable voire urgente car elle ne pouvait attendre la mise en place de la nouvelle maquette (rentrée 2017) après accréditation de la CTI qui intervient tous les six ans. 

Les visites de stage et le retour des étudiants sur leur premier emploi montrent que, formés depuis 2007 aux méthodes d'apprentissage (machine ou statistique), à la sélection de modèle et à leur utilisation concrète en projet sur données réelles, ceux-ci se sont très bien adaptés aux volume, variété, vélocité des données. Néanmoins, soucieux de l'employabilité des étudiants à venir et face à l'afflux important de telles offres de stages ou d'emplois, des contenus et formes pédagogiques ont évolué en années quatre et cinq de la spécialité à partir la rentrée 2014. 

\section{Adaptations du cursus GMM}
\subsection{Quelles compétences?}
D'un point de vue opérationnel, avant d'analyser des données il faut pouvoir les archiver, les extraire, alors que leur volume nécessite architecture et puissance de calcul inhabituelles  dans certains domaines. Dans d'autres domaines comme la recherche pétrolière ou la Météorologie, ces échelles de volume ont déjà été largement atteintes sans pour autant déclencher de tels séismes médiatiques. Puis, lorsque débute l'analyse, la "science des données" nécessite, en premier lieu, des savoir-faire de statisticien en relation avec des compétences métiers du domaine d'application. Repérer des anomalies, imputer des données manquantes, sélectionner et transformer des variables (\emph{features}), identifier des sources de variabilité, des structures de corrélation ou plus généralement de liaisons entre variables, sont des compétences basiques. Il n'est sans doute pas indispensable de toujours faire appel à un statisticien certifié, mais il est risqué de faire abstraction de ces étapes, et des compétences afférentes indispensables, avant d'aborder le travail plus sophistiqué de modélisation à nouveau statistique ou d'apprentissage qui requièrent compréhension, connaissance et expérience des méthodes utilisables. 

Un scientifique des données est finalement, idéalement, au centre d'un tétraèdre de compétences interconnectées: statistiques, informatique, mathématiques et métiers du domaine. Il maîtrise en principe une grande variété d'outils pour
\begin{itemize}
\item gérer des bases de données classiques et NOSQL (Not Only SQL), 
\item administrer des nuages de calculateurs virtuels ou non,
\item visualiser, explorer, représenter (signaux, images, graphes), sélectionner, transformer  en lien avec le métier,
\item modéliser (apprentissage statistique ou machine), prévoir tout en
\item optimisant les algorithmes, les modèles et choix de stratégie, de technologie ou plateforme. 
\end{itemize}
Il n'est évidemment pas réaliste de vouloir former en 5 ans après le bac un tel scientifique susceptible de prendre en charge toutes ces missions tout en le sensibilisant aux aspects juridiques et éthiques de confidentialité.  L'accent est mis sur la partie valorisation et donc l'analyse ({\emph{analytics}) des données massives.

\subsection{Objectifs}
L'objectif n'est pas, ou ne sera pas, de créer un nouveau Master ou une nouvelle spécialité d'ingénieur en "Sciences des Données". Le manque de ressources humaines ou de financements pour une mise en place dans le cadre de la formation continue comme c'est le cas à \href{http://www.telecom-paristech.fr/formation-continue/masteres-specialises/big-data.html}{Telecom ParisTech}, le calendrier contraint du processus d'habilitation d'une spécialité d'ingénieur, rendent irréalistes de tels objectifs. Pour les mêmes raisons budgétaires, il est impossible de démultiplier les options ou orientations. Enfin, l'appellation de ce métier "scientifique des données" est actuellement intimement liée au développement des services en ligne, pas tous viables à plus ou moins long terme et pas à l'abri d'une implosion de la bulle spéculative associée; il serait peu stratégique de négliger les métiers traditionnels pour l'avenir des étudiants et de la formation. Le choix a donc été fait d'un \emph{aménagement flexible} des contenus, au sein d'une formation professionnalisée avancée en Statistique, afin d'y introduire, au détriment d'autres aspects, les concepts et méthodes perçus, chaque année comme prioritaires pour l'analyse et la valorisation de données massives. 

Parmi toutes les compétences nécessaires à la maîtrise des principaux outils: bases de données lorsqu'elles sont massives, calcul parallèle avec données réparties, méthodes d'apprentissage machine et statistique, données textuelles ou images, problèmes spécifiques d'algorithmique, mathématique et optimisation, connaissance "métier" du domaine d'application... Le choix est donc fait d'insister sur les méthodes statistiques adaptées à l'analyse de données massives, sur les  outils mathématiques (ondelette, \emph{scatering}) et d'optimisation (convexe et stochastique) émergeants afin d'identifier les propriétés et limites de ces méthodes. En revanche certains des aspects les plus technologiques et informatiques (\emph{cloud computing}) sont abordés par leur seule mise en \oe uvre pratique afin de rendre les étudiants capable de travailler en équipe et synergie avec des spécialistes de ces outils.

Un autre objectif très important est celui de la \emph{capacité d'auto-formation}. Les technologies considérées évoluent avec une excessive volatilité. En quelques mois, certaines sont plébiscitées, d'autres sont aussi vite oubliées. Il ne s'agit donc pas de former les étudiants à des technologies dont la pérennité est très aléatoire, mais plutôt de leur apprendre à s'y former au fil de l'eau et des besoins.

\subsection{\'Evolution du cursus GMM MMS}
Les contenus des unités de formation (UF) des niveaux L3 et M1 sont progressivement aménagés. C'est relativement simple et léger dans certains cas, plus complexe dans d'autres et ne pourra se faire qu'à partir de la rentrée 2017, notamment pour l'UF centrale d'optimisation à l'interface entre modélisation numérique et stochastique.
\begin{description}
\item[Sondage] Insister (lourdement) sur la \emph{représentativité} d'un échantillon; beaucoup de don\-nées ne signifient pas beaucoup d'information ou une information suffisamment représentative des objectifs. Alors que les fichiers de connexions sur les sites sont les sources de données les plus massives et véloces, il est symptomatique de noter que \href{http://www.mediametrie.fr/mediametrie/index.php}{Médiamétrie} propose la constitution de panels d'internautes afin de certes mesurer mais surtout connaître les caractéristiques de l'audience d'un site internet.
\item[\href{http://www.math.univ-toulouse.fr/~besse/Wikistat/pdf/st-m-explo-intro.pdf}{Exploration statistique}] Ajouter la factorisation de matrices non-négatives ou NMF \href(http://wikistat.fr/pdf/st-m-explo-nmf.pdf) pour compléter le florilège des méthodes multidimensionnelles classiques basées sur une décomposition en valeurs singulières. L'illustrer par un \href{http://wikistat.fr/pdf/st-scenar-explo7-nmf.pdf}{exemple simple} de système de recommandation par filtrage collaboratif. Renvoyer au cours d'optimisation pour l'analyse détaillée des algorithmes complexes de cette méthode.
\item[Optimisation] C'est l'UF qui demande le plus de réflexion pour construire un programme avec deux objectifs: rapprocher les deux orientations "numériques" et "statistiques" de la spécialit GMM et introduire les approches innovantes: optimisation convexe avec pénalisation et exemple (travaux pratiques) d'application à la complétion de matrices, gradient stochastique avec application à venir à la régression, la SVD, la NMF... Ce programme est en préparation pour présentation à la CTI.
\item[Langages et logiciels] Les étudiants bénéficient d'un enseignement classique en Informatique (algorithmique, C, Java,  Python). Il s'agit de compléter par les langages / logiciels spécifiques au calcul et à la modélisation statistique. Une \href{http://www.math.univ-toulouse.fr/~besse/Wikistat/pdf/st-tutor1-start-R.pdf}{formation} à \href{https://cran.r-project.org/}{R} est développée car ce logiciel est de plus en plus présent dans les milieux industriels et commerciaux au détriment de \href{http://www.sas.com/fr_fr/academic.html}{SAS} dont une \href{http://www.math.univ-toulouse.fr/~besse/Wikistat/pdf/st-tutor1-sas-init.pdf}{initiation} classique reste présente mais sans introduire les services de \href{http://www.sas.com/fr_fr/software/cloud-analytics.html}{cloud analytics}. Python, devenu un standard pour le calcul scientifique est \href{http://www.math.univ-toulouse.fr/~besse/Wikistat/pdf/st-tutor1-python-start.pdf}{introduit} pour ses performances vis à vis de R, alors que les sorties tant numériques que graphiques sont encore bien pauvres. L'accent est mis sur les librairies \href{http://wikistat.fr/pdf/st-tutor2-python-pandas.pdf}{\tt pandas} pour la préparation (\emph{munging}) des données et \href{http://wikistat.fr/pdf/st-tutor3-python-scikit.pdf}{\tt scikit-learn} pour l'apprentissage. 
\item[\href{http://www.math.univ-toulouse.fr/~besse/Wikistat/pdf/st-m-app-intro.pdf}{Apprentissage statistique}] Au cours initial et en place depuis 2006 et dont le contenu est proche de celui de la référence du domaine (Hastie et al. 2002)) s'ajoutent progressivement une introduction aux techniques d'imputation de valeurs manquantes et surtout à la recherche d'atypiques (\emph{outliers}) multidimensionnels par différentes approches dont la discrimination à une classe (\emph{One Classe Classification}). Cette UF insiste sur la pratique effective sur des données réelles, même si pas massives, pour optimiser les performances (équilibre biais / variance) de ces méthodes. 
\end{description}

\subsection{Atelier}
Un \emph{Atelier: "Technologie des Sciences des Données"} est mis en place en parallèle de l'UF d'apprentissage statistique;  il s'agit d'ouvrir les accès aux principales technologies du moment dans des environnements performants.

\subsubsection*{Travail en autonomie}
Constat: un étudiant, qui a réalisé au cours de sa 4ème année un projet tutoré\footnote{12 heures encadrées sur 48 de l'emploi du temps plus 48 heures de travail personnel.} "Recherche Innovation" sur de vraies données et un stage de trois mois en entreprise en fin d'année, entre très motivé en 5ème année pour acquérir les compétences dont il a finalement compris qu'elles sont indispensables au bon début de sa carrière. Son objectif prend de l'ampleur et dépasse la seule validation de tous ses crédits ECTS. D'autre part, une priorité identifiée est de faire travailler les étudiants en équipe pluridisciplinaire afin de les faire se confronter à d'autres spécialistes pour la mise en \oe uvre de technologies très spécifiques et très mouvantes d'une année sur l'autre. 

Compte tenu de ces remarques et des objectifs visés, nous avons décidé de mettre en place des \emph{Ateliers} durant lesquels les étudiants réalisent un projet au sein d'un groupe associant des étudiants GMM virtuellement "clients", d'autres étudiants de la spécialité Informatique et Réseau (IR) et ayant choisi les mineures \emph{cloud computing}, \emph{big data} de leur spécialité. Principe: les étudiants GMM résolvent un problème de  modélisation / prévision aidés par ceux de la spécialité IR chargés de mettre en place les moyens de calcul adéquats. Ces groupes sont planifiés dans l'emploi du temps sur 40 heures avec une heure de présence d'un enseignant-chercheur par semaine et supposent autant de travail personnel par ailleurs.

Pour "aider" à la motivation les projets sont présentés sous la forme de \emph{compétitions}. Ce principe de concours a été initié depuis de nombreuses années au sein de certains colloques: \href{http://www.amstat.org/education/posterprojects/}{American Statistical Association}, \href{https://kddcup2015.com/information.html}{KDD Cup}, \href{http://chimio2015.sciencesconf.org/resource/page/id/21}{Chimiométrie}, \href{http://www.chalearn.org/challenges.html}{\it machine learning}. Il est largement popularisé par le site \href{https://www.kaggle.com/competitions}{Kaggle} et maintenant \href{https://www.datascience.net/fr/home/}{repris} par l'\href{https://www.datascience.net/fr/home/}{ENSAE ParisTech} ou l'\href{https://challengedata.ens.fr/en/home}{ENS Paris Saclay}. Il est possible, dans un premier temps d'utiliser, les capacités de la librairie {\tt rchallenge} (Todeschini et Genuer, 2015) ou de simplement demander aux étudiants de s'inscrire à un défi \href{https://www.kaggle.com/competitions}{Kaggle} en cours. Un site dédié pour l'organisation de telles compétitions entre les formations locales ou régionales sera ouvert à la rentrée 2016.

L'évaluation de cette unité de formation tient compte du classement ou de la performance de la prévision obtenue et également des temps de calcul de la phase d'apprentissage d'un modèle.

\subsubsection*{Progression pédagogique}
L'atelier progresse en parallèle d'une unité de formation approfondie (60 heures) d'\href{http://www.math.univ-toulouse.fr/~besse/Wikistat/pdf/st-m-app-intro.pdf}{appren\-tissage statistique} qui introduit les méthodes et leurs procédures d'optimisation de l'équili\-bre biais / variance. Il s'inscrit dans la continuation de la formation aux environnements de calcul statistique en suivant le fil conducteur du volume: R pour les données "petites" et des analyses complexes, Python ({\tt pandas}) pour aborder la préparation de données plus volumineuses, puis les modéliser ({\tt scikit-learn}) de façon beaucoup plus efficace qu'avec R en terme de temps de calcul. 

Le passage à l'échelle massive se base toujours sur Python en utilisant l'API \href{http://spark.apache.org/docs/latest/api/python/}{\tt pyspark} pour accéder "simplement" et efficacement à l'environnement \href{https://hadoop.apache.org/}{Hadoop}. L'accent est mis sur la librairie \href{http://spark.apache.org/mllib/}{\tt MLlib} de Spark qui exécute quelques uns des principaux algorithmes d'apprentissage non supervisés et supervisés, de factorisation (NMF par moindres carrés alternés) sur des données distribuées.

Les premières expérimentations sont faites sur les postes personnels des étudiants ou les serveurs d'enseignement de l'INSA. La librairie \href{https://github.com/RevolutionAnalytics/RHadoop}{\tt RHadoop} de R est développée par la société \href{http://revolutionanalytics.com/}{Revolution Analytics}\footnote{\href{http://blogs.microsoft.com/blog/2015/01/23/microsoft-acquire-revolution-analytics-help-customers-find-big-data-value-advanced-statistical-analysis/}{Rachetée par Microsoft} en avril 2015.} permet une initiation aux contraintes de \emph{MapReduce} de Hadoop sans avoir à installer Hadoop; l'API {\tt pyspark} permet aussi de tester les codes Python de {\tt MLlib} sans l'accès à une réelle architecture distribuée.  Dans le même ordre d'idée, \href{http://h2o.ai/}{H2O} propose aussi une librairie R dont l'utilisation permet une initiation à une version allégée d'apprentissage profond (\emph{deep learning}). 

La dernière phase de la progression consiste à transposer, tester les codes réalisés sur des \emph{clusters} commerciaux dans le cadre de partenariats (\href{https://www.ovh.com/fr/cloud/}{OVH}, \href{http://www.ibm.com/ibm/clientcenter/montpellier/}{IBM}, \href{http://www.hupi.fr/}{Hupi}). C'est à cette étape qu'interviennent les étudiants de la spécialité Informatique et Réseaux afin d'installer et optimiser les architectures. Il s'agit de mettre en correspondance performances statistiques et temps de calcul pour les différentes configurations.

\subsubsection*{Ressources}
Les ressources pédagogiques sont accessibles sur le site \href{http://wikistat.fr/}{\tt wikistat.fr} sous la forme de vignettes de présentation des \href{http://www.math.univ-toulouse.fr/~besse/Wikistat/pdf/st-m-app-intro.pdf}{éléments méthodologiques}, de tutoriels (\href{http://wikistat.fr/pdf/st-tutor2-python-pandas.pdf}{Python}, \href{http://wikistat.fr/pdf/st-tutor2-R-init.pdf}{R}) des outils logiciels et de scénarios (déroulé des commandes R ou python) d'analyse de données de plus en plus lourdes et complexes comme la \href{http://www.math.univ-toulouse.fr/~besse/Wikistat/pdf/st-scenar-app-spam.pdf}{détection de pourriels}.

Les premières séances de l'atelier sont consacrées à une auto-appropriation des outils disponibles sans faire appel à un "vrai" cluster: \href{http://www.math.univ-toulouse.fr/~besse/Wikistat/pdf/st-tutor5-R-mapreduce.pdf}{utilisation de RHadoop} pour apprécier les contraintes de \emph{MapReduce}, \href{http://www.math.univ-toulouse.fr/~besse/Wikistat/pdf/st-tutor6-R-h2o.pdf}{celle} de H20 pour introduire le \emph{deep learning} et une \href{http://www.math.univ-toulouse.fr/~besse/Wikistat/pdf/st-tutor4-pyspark-mllib.pdf}{initiation} à l'utilisation de {\tt MLlib} de Spark \emph{via} l'API {\tt pyspark}.

L'étape suivante confronte les étudiants à des données  relativement complexe et volumineuses dans la progression des outils disponibles afin d'expérimenter les limites (lenteurs) de R, la meilleure efficacité de Python, puis celle des technologies adaptées au \emph{cloud computing}. Deux situations sont proposées, d'autres seront progressivement ajoutées. 

La première concerne le problème, largement étudié dans la littérature, de reconnaissance de caractères proposé par \href{http://yann.lecun.com/exdb/mnist/}{Yan le Cun} (MINST) et base d'un \href{https://www.kaggle.com/c/digit-recognizer}{concours Kaggle} auquel les étudiants sont invités à participer. Le \href{http://wikistat.fr/pdf/st-atelier-MINST.pdf}{déroulement proposé} part de l'application naïve d'une technique de discrimination avant de tester d'autres pistes dont l'utilisation d'une \href{http://wikistat.fr/pdf/st-atelier-MINST-tangent-ditance.pdf}{distance} invariante  proposée par Simard et al. (1998).

Le deuxième sujet aborde les systèmes de recommandation par filtrage collaboratif. Après une \href{http://wikistat.fr/pdf/st-scenar-explo7-nmf.pdf}{introduction élémentaire} ce sont les données de notations du site \href{http://grouplens.org/datasets/movielens/}{movieLens} qui sont complétées par \href{http://wikistat.fr/pdf/st-atelier-recom-film.pdf}{différentes approches}.

\subsubsection*{Erreur d'optimisation}
L'un des objectifs, à ne pas perdre de vue dans la jungle technologique, est de faire évaluer expérimentalement par les étudiants le troisième terme d'erreur introduit par les contraintes imposées par les ressources. De l'emploi de R sur une simple machine, à l'utilisation de la librairie MLlib (Spark) sur un gros cluster en passant par la librairie scikit-learn de Python, les temps de calcul sont évidemment réduits de façon drastique mais l'indispensable analyse critique de ces librairies montre les restrictions qu'elles imposent en terme de choix de méthodes et de facilité d'optimisation de leur complexité.

\subsubsection*{Résultats attendus}
Les étudiants produisent des tutoriels d'installation et calepins ({\it notebook}). Ainsi, à l'issue de chaque atelier, il est facile pour les étudiants de s'approprier le travail réalisé par les autres groupes et d'élargir leur champ de compétences sur les domaines connexes. Les prévisions obtenues sur un échantillon test sont mises en compétition entre groupes et, le cas échéant, plus largement sur un site internet.

\section*{Conclusion très provisoire}
La Science des Données n'est pas une nouvelle Science mais la conséquence de nouveaux modèles économiques, de nouveaux marchés, accompagnant les nouvelles technologies. Il est certes indispensable d'y adapter la formation des étudiants  pour faciliter leur employabilité, mais aussi pour les rendre conscients des nouveaux enjeux sociaux et les responsabiliser;  c'est en marchant que nous en construisons le corpus des compétences à transmettre ou faire acquérir. Le métier de \emph{data scientist} n'est donc pas si "nouveau" et, en dépit des messages commerciaux, il est très risqué de penser qu'une valorisation optimale des données puisse être atteinte sans des compétences approfondies en modélisation statistique et apprentissage machine. Il est alors plus juste de dire qu'un statisticien est confronté à de nouveaux défis plus technologiques que méthodologiques. 

L'optimisation de la complexité d'un modèle ou le choix d'une méthode est maintenant bien balisé et relativement consensuel autour de la minimisation d'une estimation d'erreur de prévision, par exemple par validation croisée, pour équilibrer au mieux variance et biais du modèle. En revanche, le choix d'une meilleure stratégie pour minimiser la partie "erreur d'optimisation", sous contrainte de ressource de temps et de puissance de calcul, est un problème largement ouvert et ce d'autant plus que le paysage technologique est en constante (r)évolution. Les recherches sur la résolution de ces problèmes d'optimisation, leurs implémentations algorithmiques dans des architectures matérielles distribuées, de même que la prise en compte et la représentation de données complexes ou non structurées, ne constituent pas pour autant une nouvelle science. 

En définitif, le nouveau rôle du \emph{data scientist} est principalement d'associer deux types de démarches ou de logiques, celle du statisticien qui consiste à inférer, prévoir, contrôler une erreur ou un risque et celle de l'informaticien soucieux des contraintes ou performances de calcul pour atteindre l'objectif recherché.

Les objectifs pédagogiques poursuivis à l'INSA de Toulouse consistent finalement à entrainer l'apprenti statisticien à tester plusieurs stratégies dans des environnements technologiques adaptés, à développer son regard critique sur les résultats obtenus, leurs conséquences sociales, et à lui apprendre à continuer à s'auto-former pour pouvoir remettre en cause ses habitudes.

Ces objectifs sont atteints en le formant aux aspects les plus fondamentaux de la modélisation et des méthodes d'apprentissage statistique, à la représentation des données et aux problèmes d'optimisation afférents; le statisticien doit connaître, contrôler les limites des outils développés. Ces objectifs nécessitent des accès à un  éventail de technologies des données massives parmi celles disponibles du moment. Leurs cycles de développements sont nettement plus court que celui des habilitations des formations, les aménagements et adaptations doivent être très flexibles d'une année sur l'autre. On peut par exemple s'interroger sur l'impact que va avoir l'ouverture récente (novembre 2015) au public du code du projet \href{https://www.tensorflow.org/}{\it TensorFlow} de \emph{Google} qui rend plus accessible l"apprentissage profond".

Le principe de faire participer les étudiants à des concours de prévision présente des vertus pédagogiques à exploiter. Nous sommes loin d'un \emph{jeux sérieux}, façon jeu vidéo, mais il serait dommage de ne pas utiliser ce levier pour solliciter la motivation des étudiants vers  l'autonomie indispensable à l'apprentissage de l'autoformation.

La mise en place de ces moyens est en cours; il restera, avec le temps, à en évaluer la pertinence.

\section*{Références}

Bach F. et Moulines E. (2013). Non-strongly-convex smooth stochastic approximation with convergence rate O(1/n). \emph{Advances in Neural Information Processing Systems} (NIPS), Dec 2013,  \href{\tt https://hal.archives-ouvertes.fr/hal-00831977}{pdf}

Besse P., Le Gall C., Raimbault N. et Sarpy S. (2001). Data Mining \& Statistique, avec discussion, \emph{Journal de la Société Française de Statistique}, {\bf 142}, 5-35.

Besse P., Garivier A. et Loubes J.-M. (2014). Big Data Analytics - Retour vers le Futur 3 - De Statisticien à Data Scientist, \emph{Ingénierie des Systèmes d'Information}, {\bf 19/3}, 93-105.

Breiman L. (2001). Random forests. \emph{Machine Learning}, {\bf 45}, 5-32. 

Candès E. et Tao. T.  (2009). The power of convex relaxation: Near-optimal matrix completion. \emph{IEEE Trans. Inform. Theory} {\bf 56(5)}, 2053-2080. 

Cappé O., Garivier A., Maillard O. A., Munos R. et Stoltz G. (2013). Kullback-Leibler Upper Confidence Bounds for Optimal Sequential Allocation. \emph{Annals of Statistics}, {\bf 41(3}), 516-541.

Hastie T., Tibshirani R., Friedman J. (2009). \emph{The Elements of Statistical Learning: Data Mining, Inference, and Prediction}. Second Edition, Springer.

Lê Cao K.-A., Boistard S., Besse P. (2011) Sparse PLS Discriminant Analysis: biologically relevant feature selection and graphical displays for multiclass problems, \emph{ BMC Bioinformatics}, {\bf 12:253}. 

MacQueen J. (1967). Some Methods for classification and Analysis of Multivariate Observations. \emph{Proceedings of 5th Berkeley Symposium on Mathematical Statistics and Probability 1}. University of California Press, 281-297.

Simard P., Le Cun Y., Denker J. et  Victorri B. (1998). Transformation invariance in pattern  recognition - tangent distance and tangent propagation, in \emph{Neural Networks: Tricks of the Trade}, second edition, Lecture Notes in Computer Science, Springer.

Todeschini A. et Genuer R. (2015)  Compétitions d'apprentissage automatique avec le package R  rchallenge.   47èmes  Journées  de  Statistique  de  la  SFdS,  Juin  2015,  Lille,  France.   \href{https://hal.inria.fr/hal-01157147}{pdf}.

Zaharia M., Chowdhury M., Das T. and Dave A., Ma J. McCauly M., Franklin M.,  Shenker S. et Stoica I. (2012). Resilient Distributed Datasets: A Fault-Tolerant Abstraction for In-Memory Cluster Computing, in {\it USENIX Symposium on Networked Systems Design and Implementation}  (NSDI 12), 15-28.

\end{document}